\documentclass[%
 reprint,
 amsmath,amssymb,
 aps,
pra,
floatfix,
]{revtex4-1}
\usepackage{graphicx,lipsum}
\usepackage{dcolumn}
\usepackage{bm}
\usepackage[dvipsnames]{xcolor}
\usepackage{array,multirow}
\usepackage{bm}
\usepackage{bbm}
\usepackage{amsmath}

\begin{document}

\preprint{APS/123-QED}
\title {Quantum speed limit for the creation and decay of quantum correlations}

\author{K.G. Paulson}
\altaffiliation[paulsonkgeorg@gmail.com]{}
\author{Subhashish Banerjee }
\email{subhashish@iij.ac.in }
\affiliation{Indian Institute of Technology, Jodhpur-342030, India}



\date{\today}

\begin{abstract}
We derive Margolus-Levitin and Mandelstamm-Tamm type bound on the quantum speed limit  time for the creation and decay of quantum correlations by an amount in a quantum system evolving under the influence of its ambient environment. The minimum distance of a non-classical state from an appropriate set of classical states is a legitimate measure of the quantumness of the state. We consider entanglement and quantum discord measures of quantum correlations, quantified using the Bures distance-based measure. To demonstrate the impact of quantum noise on this speed limit time for quantum correlations, we estimate the quantum speed limit time for the creation and decay of quantum correlations for a two-qubit system under modified OUN dephasing and collective two-qubit decoherence channels.

\end{abstract}

\maketitle


\section{\label{sec:level1} Introduction }  
The investigation of quantum speed limit (QSL) time for the  evolution of quantum states  discusses  the fundamental problem of the rate at which quantum states and their characteristics change. Since the  interpretation of energy-time uncertainty principle as the limit on the time of evolution in physical process, a dedicated amount of work has been done to  report the significant applications of quantum speed limit in the field of quantum information processing and associated domains~\cite{deffner2017quantum}. Mandelstam, Tamm (MT)~\cite{mandelstam1945} and  Margolus, Levitin  (ML)~\cite{margolus1998} derived bounds on the minimum time needed for a quantum system to evolve between the orthogonal states for closed systems. Similarly, in~\cite{deffner2013energy} QSL time for achieving a target fidelity for driven quantum systems has been derived. The  fact that the impact of the bath on a quantum system is not always detrimental in nature has sped up the research work unraveling  the characteristics of open quantum systems~\cite{sbbook}, and naturally the investigation of speed of evolution under non-unitary dynamics has become an active research topic~\cite{murphy2010communication,paulson2021hierarchy,paulson2021effect,paulson2022quantum,baruah2022phase}. Quantum speed limit for open quantum system based on different measures has been proposed~\cite{deffner2013quantum,del2013quantum,taddei2013quantum}. The investigation of QSL time realizes wide range of applications in the field of quantum information processing and technology as it reflects the nature of the physical process the quantum system undergoes~\cite{paulson2021effect}.  Thus for example, the memory effects on the dynamics of quantum systems originating from the  coupling  between the system and bath results in the decay-revival mechanism of quantum correlations, which is very well captured by the quantum speed limit time for certain physical processes~\cite{paulson2021hierarchy}.\newline
The existence of quantum correlations and coherence in a system is a valuable resource for many tasks of fundamental and practical importance. The uncontested supremacy of certain protocols in quantum regime over classical counterparts in the field of quantum metrology, quantum communication, quantum thermodynamics,~\cite{paulson2019tripartite, sapienza2019correlations} etc., is achieved   by manipulating and controlling quantum correlations and coherence. The  interwoven connection between the coherence and quantum entanglement~\cite{streltsov2015measuring,bhattacharya2018evolution} signifies the need of harnessing their stronger interdependence and the fundamental speed limit at which  a system reveals its quantumness. Recent works in this direction reveal that   speed limit is an inherent feature of system's dynamics in the Hilbert space ~\cite{okuyama2018quantum,shanahan2018quantum}. 

The significance of quantum correlations as a resource, and technique to generate and manipulate them seeks extensive consideration in the branch of quantum communication and technology. A number of techniques have been proposed for creating entanglement in quantum systems~\cite{schneider2002entanglement,kim2002entanglement,tanas2004entangling}, which is  one of the  themes of the present work.  \newline
In this work, we address the question  how fast quantum correlations can be processed in a system under a physical process. To this end, we derive  bounds on the minimum time required for the generation and decay of quantum correlations by an amount in a quantum system for non-unitary evolution. We use distance based  measures of quantum entanglement and discord to derive the QSL time valid for open quantum systems. We  use the  Bures metric (geometric measure)   measures to estimate quantum entanglement and discord.  We illustrate our ideas, for the generation and decay of quantum correlations, on two models, {\it viz.}, the modified Ornstein–Uhlenbeck noise (OUN), which is dephasing, and a two two-level atomic dipole-dipole interaction model. We discuss the conditions under which quantum correlations are  created in an initial separable atomic system.   We estimate the QSL time for the change in  quantum correlations in the atomic system, and analyze how  the coupling strength impacts the process.
\newline
The present work is organized as follows. In Sec. II, we discuss the preliminaries required for the current work. The derivation of the minimum time required for the generation and decay of the quantum correlations under non-unitary evolution is given in Sec. III. In Sec. IV, we consider the example of  a local dephasing channel and  a two two-level atomic system, and investigate the QSL time for the creation and decay of quantum correlations. The concluding remarks are laid down in Sec. V. 
\section{Preliminaries}

\subsection{Distance measure of quantum correlations}

\subsubsection{Measures of entanglement}
In the literature, different methods are  used to quantify entanglement. In principle a good measure of entanglement $E(\rho)$ has to satisfy the following conditions~\cite{vedral1997quantifying}. i) $E(\rho)=0$ for separable state $\rho$, ii) Local operation leaves the entanglement invariant, i.e., $E(\rho)= E(U_{A}\otimes U_{B} \rho U_{A}^{\dag } \otimes U_{B}^{\dag})$, iii) Local operation and classical communication (LOCC) cannot increase the entanglement, i.e., $E(\textrm{LOCC }\rho)\leq E(\rho)$. Let us consider a set $\mathcal{S}$ of all quantum states, is divided into two subsets  of entangled $S_{E}$ states and product (separable) $S_{P}$  states, respectively. Along with a set of classical states $S_{C}$ it shows $S_C\subseteq S_P\subseteq S_{E}\subseteq\mathcal{S}$. The distance measure of entanglement of quantum state  $\rho$ is expressed as
\begin{equation}
    E(\rho)=\min_{\sigma\in S_{P}} \chi(\rho\vert\vert\sigma),
    \label{ent_measure}
\end{equation}
where $\chi$ can be any measure of distance between the density matrices. The Bures metric $\chi_{B}$ as a measure of entanglement,
\begin{equation}
    E^{\mathcal{B}}(\rho)=\chi_{\mathcal{B}}(\rho\vert\vert\sigma)=\frac{1}{2}d_{B}^2(\rho,\sigma)=\max_{\sigma\in S_p}(1-\sqrt{F_P(\rho)}),
    \label{ent_bures}
\end{equation}
where $F_{P}(\rho)$ is the maximum fidelity between the entangled state $\rho$ and a product state in $ S_{P}$. To calculate $E^{\mathcal{B}}(\rho)$, instead of the minimization in Eq.~\ref{ent_measure}, we maximize $F_{P}(\rho)$ (Eq.~\ref{ent_bures}),  fidelity between $\rho$  and a separable state in $S_{P}$. The fidelity is the  squared overlap of quantum states 
$F(\rho,\sigma)=\bigg[\textrm{Tr}[\sqrt{\sqrt{\sigma}\rho\sqrt{\sigma}}]\bigg]^{2}$.

\subsubsection{Quantum Discord}
Consider a bipartite quantum system $A B$ with Hilbert space $H=H_{A}\otimes H_{B}$,  $N_{A}$ and $N_{B}$ arbitrary dimensions of subsystems $A$ and $B$ with spaces $H_{A}$ and $H_{B}$. The reduced subsystems $\rho_{A,B}=\textrm{Tr}_{B,A}(\rho)$. The total correlations of the composite system is given by the mutual information $I_{A:B}(\rho)=S(\rho_A)+S(\rho_B)-S(\rho)$, where $S(\rho)=-\sum_i\lambda_i \log\lambda_i$ is the von Neumann entropy. A classically equivalent definition of mutual information is defined $C_{A}(\rho)=S(\rho_{B})-\min_{E_{k}}\sum_{k}p_{k}S(\rho_{B\vert k})$, where $\rho_{B\vert k}=\textrm{Tr}_A(E_k\otimes I_B \rho)/\textrm{Tr}(E_k\otimes I_{B}\rho)$ is the state of $B$ conditioned on outcome $k$ in $A$, $\{E_{k}\}$ describes the set of POVM elements. Quantum discord is defined as the difference between the two measures of information
\begin{equation}
    D_{A}(\rho)=I(\rho)-C_{A}(\rho).
    \label{discord_def}
\end{equation}
The quantum discord is zero for classical states $(D_{A}=D_{B}=0$), otherwise is always non-negative and  $D_{A}\neq D_{B}$. The quantum discord  given in Eq.~\ref{discord_def}  requires extensive numerical minimization. In~\cite{dakic2010necessary} a geometrical measure of  quantum discord $    D_{A}(\rho)=\min_{\sigma\in S_{C}}\vert\vert\rho-\sigma\vert\vert^2
$  has been proposed, where $S_C$ is a set of classical states.
\begin{equation}
    D_{A}(\rho)=\min_{\sigma\in S_{C}}\vert\vert\rho-\sigma\vert\vert^2,
    \label{discord_gm}
\end{equation}
$\vert\vert A-B\vert\vert^2=\textrm{tr}(A-B)^2$. The geometrical discord was used to provide, among others, an operational meaning to quantum teleportation~\cite{adhikari2012operational}.
The Bures metric for quantum discord ~\cite{spehner2013geometric} is 
\begin{equation}
    D^{\mathcal{B}}_{A}(\rho)=\chi_{\mathcal{B}}(\rho\vert\vert\sigma)=\frac{1}{2}d_{B}^2(\rho,\sigma)=\max_{\sigma\in S_{C}}(1-\sqrt{F_C(\rho)}),
    \label{discord_bures}
\end{equation}
where $F_C(\rho)$ is the maximum fidelity between $\rho$ and a classical state in $S_{C}$.

\section{Minimum time for the change of quantum correlations in system}
Here, we derive the minimum time required to witness a change in quantum correlations $(Q{C})$ by an amount $\mathcal{Q}(\rho)=\vert\mathcal{Q}_{C}(\rho_0)-\mathcal{Q}_{C}(\rho_{t})\vert$ for a system evolving from initial state to final state under non-unitary dynamics. To estimate the QSL time for the change of QC, which we call ${\tau_{QC}}$, the distance between the initial and final states under evolution is calculated as a measure to quantify the change in  quantumness of the state. In this work, we mainly consider entanglement and quantum discord as signatures of quantumness of the states. Bures distance based measures are made use of to estimate the quantum correlations.
 \subsection{Bures metric as a measure of quantum correlations}
We define the quantum speed limit time for the creation or decay  of entanglement by an amount $E^{\mathcal{B}}(\rho)$ in a system for  a non-unitary evolution of the form
\begin{equation}
    \dot{\rho_t}=L_t\rho_t,
\end{equation}
where, $L_t$, the generator of the dynamics. To begin with we rewrite Eq.~\ref{ent_bures} 
\begin{equation}
E^{\mathcal{B}}(\rho)=\frac{1}{2}d_{B}^2(\rho,\sigma)=\max_{\sigma\in S_p}(1-\sqrt{F_P(\rho)}),
\label{bures_puri}
\end{equation}
where,
\begin{equation}
    F_P(\rho)=F(\rho,\sigma)=\max\vert\langle\psi_{\rho}\vert\phi_{\sigma}\rangle\vert^2,
\end{equation}
is the Uhlmann fidelity~\cite{jozsa1994fidelity}. Here $\vert\psi_{\rho}\rangle$ is the purification of the state $\rho$ and the maximization is done over all  purification $\vert\phi_\sigma\rangle$ of separable state $\sigma$ in the extended Hilbert space $H_{\mathcal{A}}\otimes H_{\mathcal{B}}$. The states $\rho$ and $\sigma$ are the reduction of these pure states over the ancillary Hilbert space $H_{\mathcal{B}}$: $\rho=\textrm{Tr}_{\mathcal{B}}\vert\psi_\rho\rangle\langle\psi_\rho\vert$ and $\sigma=\textrm{Tr}_{\mathcal{B}}\vert\phi_\sigma\rangle\langle\phi_\sigma\vert$.\newline
We define the variation in entanglement observed on the system's evolution between initial state $\rho_0$ and final state $\rho_t$ as
\begin{equation}
\small
   E^{\mathcal{B}}(\rho)=\vert E^{\mathcal{B}}(\rho_0)-E^{\mathcal{B}}(\rho_t)\vert,
   \label{ent_rho}
 \end{equation}
equivalently 
\begin{equation}
E^{\mathcal{B}}(\rho)= 
\begin{cases}
    E^{\mathcal{B}}(\rho_0)-E^{\mathcal{B}}(\rho_t),& \text{if } E^{\mathcal{B}}(\rho_0)\geq E^{\mathcal{B}}(\rho_t)\\~\\
    E^{\mathcal{B}}(\rho_t)-E^{\mathcal{B}}(\rho_0), & \text{otherwise}.
\end{cases}
\label{ent_rho_1}
\end{equation}

Hereafter, $E^{\mathcal{B}}(\rho)$, as in Eq.~\ref{ent_rho_1}, is called the change in entanglement occurred due to the decay or creation of entanglement in the system,  unless otherwise mentioned.  By substituting Eq.~\ref{bures_puri} in the latter equation we get

\begin{equation}
E^{\mathcal{B}}(\rho)= 
\begin{cases}
    -\sqrt{F_P(\rho_0)}+\sqrt{F_P(\rho_t)},& \text{if } E^{\mathcal{B}}(\rho_0)\geq E^{\mathcal{B}}(\rho_t)\\~\\
    \sqrt{F_P(\rho_0)}-\sqrt{F_P(\rho_t)}, & \text{otherwise}.
\end{cases}
\label{ent_rho_11}
\end{equation}

We take the temporal rate of change of entanglement $E^{\mathcal{B}}(\rho)$ 
 \begin{multline}
  \frac{d}{dt} E^{\mathcal{B}}(\rho)\leq   \vert \frac{d}{dt} E^{\mathcal{B}}(\rho)\vert=\frac{\vert\dot{F_P}(\rho_t)\vert}{2\sqrt{F_P(\rho_t)}}\Rightarrow\\ \frac{d{E}^{\mathcal{B}}(\rho)}{dt}\leq\frac{\vert\langle\phi_\sigma\vert\dot{\rho}_{\psi}(t)\vert\phi_\sigma\rangle+\langle\psi_\rho\vert\dot{\sigma_{\phi}}(t)\vert\psi_\rho\rangle\vert}{2(1-E^{\mathcal{B}}(\rho_t))},
 \end{multline}
where $\rho_{\psi}(t)=\vert\psi_\rho(t)\rangle\langle\psi_\rho(t)\vert$, $\sigma_{\phi}(t)=\vert\phi_\sigma(t)\rangle\langle\phi_\sigma(t)\vert$ are the purification of the states $\rho$ and $\sigma$, respectively. By substituting $E^{\mathcal{B}}(\rho_t)$ from Eq.~\ref{ent_rho_1}, we write
\begin{align}
\small
\nonumber
    2(1-(E^{\mathcal{B}}(\rho_0)\mp E^{\mathcal{B}}(\rho)))\frac{dE^{\mathcal{B}}(\rho)}{dt}
    \leq\big\vert\langle\phi_\sigma\vert\dot{\rho}_{\psi}(t)\vert\phi_\sigma\rangle+\\\langle\psi_\rho\vert\dot{\sigma_{\phi}}(t)\vert\psi_\rho\rangle\big\vert
    =\vert\langle\phi_\sigma\vert L_t{\rho}_{\psi}(t)\vert\phi_\sigma\rangle+\langle \psi_\rho\vert L_t{\sigma}_{\phi}(t)\vert\psi_\rho\rangle\vert,
    \label{esl_time}
\end{align}
for the non-unitary evolution, where $\mp$ sign indicates the decay and creation of entanglement, respectively. We begin by providing a derivation for Margolus-Levitin bound. With $\langle\phi_\sigma\vert L{\rho}_{\psi}(t)\vert\phi_\sigma\rangle=\textrm{tr}\{\vert\phi_\sigma\rangle\langle\phi_\sigma\vert L_t\rho_{\psi}(t)\}$, and $\langle\psi_\rho\vert L_t{\sigma}_{\phi}(t)\vert\psi_\rho\rangle=\textrm{tr}\{\vert\psi_\rho\rangle\langle\psi_\rho\vert L_t\sigma_{\phi}(t)\}$, Eq.~\ref{esl_time} is rewritten as
\begin{align}
\nonumber
    2(1-(E^{\mathcal{B}}(\rho_0)\mp E^{\mathcal{B}}(\rho)))\frac{dE^{\mathcal{B}}(\rho)}{dt}\\\leq\vert\textrm{tr}\{\sigma_{\phi}(t) L_t{\rho}_{\psi}(t)\}+\textrm{tr}\{\rho_{\psi}(t) L_t{\sigma}_{\phi}(t)\}\vert.
    \label{esl_time_2}
\end{align}
The
 von-Neumann trace  inequality for operators states that
\begin{equation}
    \textrm{tr}\vert B_1 B_2\vert\leq\sum_{i=1}^n\mu_{1,i}\mu_{2,i},
    \label{singular_val}
\end{equation}
where $\mu_{1,i}$ and $\mu_{2,i}$ are the singular values in the descending order for any complex matrices $B_1$ and $B_2$, respectively. For a Hermitian operator the absolute value of the eigenvalue gives the singular value. By combining Eqs.~\ref{esl_time_2} and ~\ref{singular_val} and using triangle inequality for absolute values $(\vert A+B\vert\leq \vert A\vert+\vert B\vert)$ we get
\begin{equation}
\begin{split}
\small
 2(1-(E^{\mathcal{B}}(\rho_0)\mp E^{\mathcal{B}}(\rho)))\dot{E}^{\mathcal{B}}(\rho)\\
\leq\sum_i\mu_{1,i} q_{1,i}+\sum_i\mu_{2,i} q_{2,i}=\mu_{1,1}+\mu_{2,1}
      \label{ent_bures_qsl_op}
      \end{split}
\end{equation}
where $\mu_{1,i}$ $\mu_{2,i}$ are the singular values of $L_t \rho_{\psi}(t)$ and $L_t \sigma_{\phi}(t)$, respectively. We have $q_{1,1}=q_{2,1}=1$ for pure state $\sigma_{\phi}(t)$ and $\rho_{\psi}(t)$. The largest singular value of a Hermitian operator gives the operator norm $\vert\vert L_t (\rho_\psi (t))\vert\vert_{\textrm{op}}=\mu_{1,1}$ and $\vert\vert L_t (\sigma_\phi (t))\vert\vert_{\textrm{op}}=\mu_{2,1}$. Equation~\ref{ent_bures_qsl_op} can be rewritten using von-Neumann inequality condition as
\begin{equation}
\begin{split}
\small
      2(1-(E^{\mathcal{B}}(\rho_0)\mp E^{\mathcal{B}}(\rho)))\dot{E}^{\mathcal{B}}(\rho)\leq  \vert\vert L_t (\rho_\psi (t))\vert\vert_{\textrm{op}}+\vert\vert L_t (\sigma_\phi (t))\vert\vert_{\textrm{op}}\\
      \leq \vert\vert L_t (\rho_\psi (t))\vert\vert_{\textrm{tr}}+\vert\vert L_t (\sigma_\phi (t))\vert\vert_{\textrm{tr}},
      \label{ent_bures_qsl}
      \end{split}
\end{equation}

and integrating over  time we get
\begin{equation}
\small
    \tau\geq\tau_{QC}=\max\Bigg\{\frac{1}{K_{\textrm{op}}^{\tau}},\frac{1}{K_{\textrm{tr}}^{\tau}}\Bigg\} 2 E^{\mathcal{B}}(\rho)\bigg(1-\frac{2 E^\mathcal{B}(\rho_0)\mp E^{\mathcal{B}}(\rho)}{2}\bigg),
    \label{E_spdlmt_op_tr}
\end{equation}
where we write
$K_{\textrm{op,tr}}^{\tau}=1/\tau\int_0^{\tau}dt(\vert\vert L_t{\rho_{\psi}(t)}\vert\vert_{\textrm{op,tr}}$+ $\vert\vert L_t{\sigma_{\phi}(t)}\vert\vert_{\textrm{op,tr}})$.\newline
To derive Mandelstamm-Tamm bound we rewrite Eq.~\ref{esl_time} as
\begin{equation}
\begin{split}
    2(1- (E^{\mathcal{B}}(\rho_0)\mp E^{\mathcal{B}}(\rho))\dot{E}^{\mathcal{B}}(\rho)\leq \vert\textrm{tr}\{ L_t{\rho}_{\psi}(t)\vert\phi_\sigma\rangle\langle\phi_\sigma\vert\}\vert\\+\vert\textrm{tr}\{ L_t{\sigma}_{\phi}(t)\vert\psi_\rho\rangle\langle\psi_\rho\vert\}\vert.
    \end{split}
\end{equation}
Using Cauchy-Schwarz inequality for operators, above equation takes the form
\begin{equation}
\begin{split}  
\nonumber
2(1- (E^{\mathcal{B}}(\rho_0)\mp E^{\mathcal{B}}(\rho)))\dot{E}^{\mathcal{B}}(\rho)\leq  \\ \sqrt{\textrm{tr}\{  L_t{\rho}_{\psi}(t)L_t{\rho}_{\psi}(t)^{\dag}\}\textrm{tr}\{( \vert\phi_\sigma\rangle\langle\phi_\sigma\vert)^2\}} \\ +\sqrt{\textrm{tr}\{ L_t{\sigma}_{\phi}(t)L_t{\sigma}_{\phi}(t)^{\dag}\}\textrm{tr}\{( \vert\psi_\rho\rangle\langle\psi_\rho\vert)^2\}},
\end{split}
\end{equation}
 with $\textrm{tr}\{\vert\phi_\sigma\rangle\langle\phi_\sigma\vert^2\}=1=\textrm{tr}\{\vert\psi_\rho\rangle\langle\psi_\rho\vert^2\}$ for pure states $\vert\phi_\sigma\rangle$ and $\vert\psi_\rho\rangle$, and it gives,
 \begin{equation}
 \begin{split}
    2(1-( E^{\mathcal{B}}(\rho_0)\mp E^{\mathcal{B}}(\rho)))\dot{E}^{\mathcal{B}}(\rho)\leq  \sqrt{\textrm{tr}\{ L_t{\rho}_{\psi}(t)L_t{\rho}_{\psi}(t)^{\dag}\}}\\+\sqrt{\textrm{tr}\{ L_t{\sigma}_{\phi}(t)L_t{\sigma}_{\phi}(t)^{\dag}\}}=\vert\vert L_t\rho_\psi\vert\vert_{\textrm{hs}}+\vert\vert L_t\sigma_\phi\vert\vert_{\textrm{hs}},
    \end{split}
 \end{equation}
where $\vert\vert B\vert\vert_{\textrm{hs}}=\sqrt{\textrm{tr} \{B B^{\dag}\}}=\sqrt{\sum_i\mu_i^2}$ is the Hilbert-Schmidt
norm. Integrating the latter over time gives MT type bound for achieving a target quantum correlation under non-unitary dynamics,

\begin{equation}
    \tau\geq\tau_{QC}=\frac{1}{K_{\textrm{hs}}^{\tau}}2 E^{\mathcal{B}}(\rho)\bigg(1-\frac{ 2 E^\mathcal{B}(\rho_0)\mp E^{\mathcal{B}}(\rho)}{2}\bigg),
    \label{E_spdlmt_hs}
\end{equation}
where $K_{\textrm{hs}}^{\tau}=1/\tau\int_0^{\tau}dt(\vert\vert L_t\rho_\psi\vert\vert_{\textrm{hs}}+\vert\vert L_t\sigma_\phi\vert\vert_{\textrm{hs}})$.



Combining Eqs.~\ref{E_spdlmt_op_tr} and ~\ref{E_spdlmt_hs} we obtain,
\begin{equation}
\small
\tau_{QC}=\max\Bigg\{\frac{1}{K_{\textrm{op}}^{\tau}},\frac{1}{K_{\textrm{tr}}^{\tau}},\frac{1}{K_{\textrm{hs}}^{\tau}}\Bigg\} 2 E^{\mathcal{B}}(\rho)\bigg(1-\frac{ 2 E^\mathcal{B}(\rho_0)\mp E^{\mathcal{B}}(\rho)}{2}\bigg),
    \label{E_spdlmt}
\end{equation}
where we have $K_{\textrm{op,tr,hs}}^{\tau}=1/\tau\int_0^{\tau}dt( \vert\vert L_t\rho_\psi\vert\vert_{\textrm{op,tr,hs}}+\vert\vert L_t\sigma_\phi\vert\vert_{\textrm{op,tr,hs}})$. Equation~\ref{E_spdlmt} provides a unified MT-ML bound on the minimum time required to observe a change in quantum  entanglement by an amount $E^{\mathcal{B}}(\rho)$ in an open quantum system for its evolution from an initial pure state to a final state.\newline

Similarly, the  minimum time for the change of quantum discord by an amount $D^{\mathcal{B}}_A(\rho)$ in a system under non-unitary dynamics is given by
\begin{equation}
\small
    \tau_{QC}=\max\Bigg\{\frac{1}{K_{\textrm{op}}^{\tau}},\frac{1}{K_{\textrm{tr}}^{\tau}},\frac{1}{K_{\textrm{hs}}^{\tau}}\Bigg\} 2 D^{\mathcal{B}}_A(\rho)\bigg(1-\frac{ 2D^{\mathcal{B}}_A(\rho_0)\mp D^{\mathcal{B}}_A(\rho)}{2}\bigg),
    \label{D_spdlmt}
\end{equation}
where $D^{\mathcal{B}}_A(\rho)= \vert D_A^{\mathcal{B}}(\rho_{0})-D_A^{\mathcal{B}}(\rho_t)\vert$, is calculated using Eq.~\ref{discord_bures}, and  $K_{\textrm{op,tr,hs}}^{\tau}$ takes the form as  that in Eq.~\ref{E_spdlmt}.

\section{Models}
To demonstrate the derived speed limit time bound on the decay and creation of quantum correlations, we  consider two models of decoherence channels. First channel is the modified Ornstein–Uhlenbeck noise (OUN), which is purely a dephasing channel, and the second one is a collective two-qubit decoherence model.
\subsection{Modified Ornstein–Uhlenbeck noise (OUN)}
\begin{figure}[!ht]
    \centering
    \includegraphics[height=65mm,width=1\columnwidth]{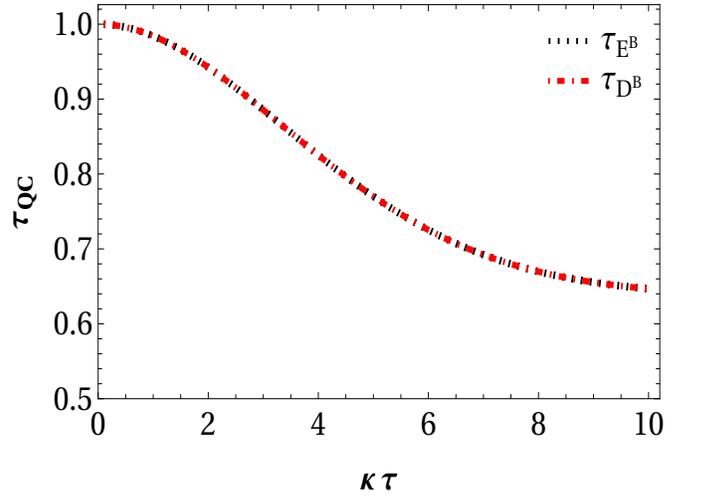}
    \caption{Quantum speed limit time in terms of Bures distance for the decay of quantum correlations  by an amount $(E^{\mathcal{B}}(\rho)$, $D_A(\rho))$ for a maximally entangled Bell state $\vert\psi^{+}\rangle$  evolved under  OUN channel.The coupling parameter is $\lambda=0.1\kappa$.  The actual driving time $\tau=1$.}
    \label{qslt_bures_ed_decay}
\end{figure}
The dynamics of a quantum system under dephasing, a unital process~\cite{yu2010, Shrikant2018, Pradeep,shrikant2020} is given by the master equation,
\begin{equation}
    \dot{\rho_{t}}=\gamma(t)(s_{z}\rho_{t} s_{z}-\rho_{t}).
    \label{depmas}
\end{equation}
The decoherence function of OUN is
\begin{equation}
    p_t=  e^{\frac{-\kappa}{2}\{t+\frac{1}{\lambda}(e^{-\lambda t}-1)\}},
    \label{OUNfn}
\end{equation}
where, $\lambda^{-1}\approx\tau_{r}$ defines reservoir's finite correlation time and $\kappa$ is the coupling strength related to qubit's relaxation time.
The decoherence rate $\gamma(t)=-\frac{\dot{p_{t}}}{2 p_t}$ is calculated as
\begin{equation}
    \gamma(t) =
          \frac{\kappa (1-e^{-\lambda t})}{4}.
\end{equation}

It is noteworthy that calculating  Bures distance gives the quantification of quantum correlations and requires the knowledge of the nearest classical states. Identifying the closest classical states makes estimating the Bures measure for quantum correlation a tricky task. Here, we use the analytical expressions of Bures distance for entanglement and discord  to compute QSL time for QC.  For two-qubit states, the fidelity of separability as a function of concurrence as a measure of entanglement~\cite{vidal2000optimal, wei2003geometric,streltsov2010linking} is given as
\begin{equation}
    F_p(\rho)=\max_{\sigma\in S_p}F(\rho,\sigma)=\frac{1}{2}(1+\sqrt{1-C(\rho)^2}),
    \label{Bures_ent_fidelity}
\end{equation}
where, $C(\rho)=\max\{0,\kappa_1-\kappa_2-\kappa_3-\kappa_4\}$, $\kappa_i^{'s}$ are the eigenvalues of the matrix $\sqrt{\sqrt{\rho}\tilde{\rho}\sqrt{\rho}}$ in the descending order, with $\tilde{\rho}=\sigma_y\otimes\sigma_y\rho^*\sigma_y\otimes\sigma_y$, $\rho^*$ is the complex conjugate of the density matrix $\rho$. Substituting $F_p(\rho)$ in Eq.~\ref{ent_bures}, we get  Bures measure of entanglement 
\begin{equation}
    E^{\mathcal{B}}(\rho)=(1-\sqrt{\frac{1+\sqrt{1-C(\rho^2)}}{2})}.
\end{equation}
For  maximally entangled Bell states $E^{\mathcal{B}}(\rho)=1-1/\sqrt{2}$.\newline

Similarly, we calculate QSL time for the change of quantum discord in terms of Bures measure (Eq.~\ref{discord_bures}). For a pure state $(\rho=\vert\psi\rangle\langle\psi\vert)$, it can be shown that $D^{\mathcal{B}}_{A}(\rho)=D^{\mathcal{B}}_{B}(\rho)=E^{\mathcal{B}}(\rho)=2(1-\sqrt{\mu_{\max}})$, where $\mu_{\max}$  is the highest eigenvalue of the state $\rho$. The equality between the Bures measure of  discord and entanglement occurs from the fact that the closest product state to a pure entangled state is a pure separable state. For a two-qubit Bell diagonal state $\rho_B=\sum_i p_i\vert B_i\rangle\langle B_i\vert$, ($\{B_i\}$ is the maximally entangled basis set), the  measure of Bures discord~\cite{spehner2013twoqubit} is given by

\begin{equation}
   D^{\mathcal{B}}_{A}(\rho)=2\bigg(1-\sqrt{F_{C}(\rho)}\bigg)= 2\Bigg(1-\sqrt{\frac{1+b_{\max}}{2}}\Bigg),
   \label{Bures_dsc_fidelity}
\end{equation}
 where, 
 \begin{equation}
 \small
\begin{split}
\nonumber
b_{\max}=\frac{1}{2}\max\{\sqrt{(1+c_1)^2-(c_2-c_3)^2}
+\sqrt{(1-c_1)^2-(c_2+c_3)^2},\\\nonumber
\sqrt{(1+c_2)^2-(c_1-c_3)^2}+\sqrt{(1-c_2)^2-(c_1+c_3)^2},\\
\sqrt{(1+c_3)^2-(c_1-c_2)^2}+\sqrt{(1-c_3)^2-(c_1+c_2)^2}\},
\end{split}
\end{equation}
and $c_i=\textrm{tr}(\rho \sigma_i\otimes\sigma_i)$ for $i=1,3$. For a maximally entangled Bell state $D^{A}(\rho)=1-1/\sqrt{2}$. \newline
For a maximally entangled Bell state $\vert\psi^{+}\rangle$ under local OUN noise, using Eqs.~\ref{Bures_ent_fidelity} and \ref{Bures_dsc_fidelity}, we calculate $F_{P}(\rho)=F_{C}(\rho)=1/2(1+\sqrt{1-\exp(-2\kappa( t+\frac{-1+e^{-t\lambda}}{\lambda})}))$, which implies the same expressions for Bures measures of entanglement and discord.  This ensures that the closest separable and classical states to maximize the overlap between the correlated state are equivalent. For the maximally entangled state considered, separable state $\sigma=\frac{1}{2}(\vert01\rangle\langle01\vert+\vert10\rangle\langle10\vert)$ gives the maximum fidelity. The time dependent state evolved under  OUN dephasing noise maintains its Bell diagonal form, and its overlap with the same separable state $\sigma$  provides the maximum fidelity.\newline
In Fig.~\ref{qslt_bures_ed_decay}, ML-type bound for the speed limit time $(\tau_{E^{\mathcal{B}}},\tau_{D^\mathcal{B}})$ for the decay of quantum entanglement and discord for maximally entangled Bell state $(\vert\psi^{+}\rangle)$ in the case of OUN noise is given as a function of coupling strength. We see that even though the speed limit time for the decay of quantum correlations increases  in small time intervals, overall it decreases as the coupling strength increases.\newline

\subsection{Collective two-qubit decoherence model}
To discuss the quantum speed limit for both the creation and decay of quantum correlations,  we consider a system of two two-level atoms, which have ground and excited states $\vert g_{i}\rangle, \vert e_{i}\rangle$ $(i=1,2)$, connected to a vacuum bath by  dipole transition moments $\vec{d_{i}}$. Two atoms coupled to all modes of the EM  field in vacuum, are located at the positions $\vec{r_{1}}$ and $\vec{r_{2}}$, respectively. The master equation for the time evolution of the atomic system coupled through the vacuum field is ~\cite{ficek2002entangled, banerjee2010dynamics,banerjee2010entanglement,Indranil2011}
\begin{multline}
    \frac{\partial \rho}{\partial  t}=-i \sum_{i=1}^2\omega_i[S_i^2,\rho]-i\sum_{i\neq j}M_{i,j}[S_i^{+}S_{j}^{-},\rho]\\
    -\frac{1}{2}\sum_{i,j=1}^2 \Lambda_{i,j}(\rho S_{i}^{+}S_{j}^{+}+S_{i}^{+}S_{j}^{-}\rho-2S_{j}^{-}\rho S_{i}^{+}),
    \label{mast_eq}
\end{multline}
where $S^{+}_{i}$ ($S^{-}_{i}$) are the dipole raising (lowering) operators and $S^{z}_{i}$ the energy operator of the $i$th atom. $\Lambda_{ii}$ are the spontaneous emission rates of the atoms. $\Lambda_{ij}$  are the collective spontaneous emission rates arising from the coupling between the atoms through the
vacuum bath and  $M_{ij}$ $(i\neq j)$ (dipole-dipole) represent the inter-atomic coupling. The collective damping rate $\Lambda_{ij}$ and dipole-dipole interaction potential $M_{ij}$ are defined 
\begin{multline}
    \Lambda_{ij}=\Lambda_{ji}=\frac{3}{2}\sqrt{\Lambda_i\Lambda_j}\bigg\{\bigg[1-(\vec{d}.\vec{r}_{ij})^2\bigg]\frac{\sin(\mu_{0}r_{ij})}{\mu_{0}r_{ij}}\\
    +\bigg[1-3(\vec{d}.\vec{r}_{ij})^2\bigg]\bigg[\frac{\cos(\mu_0 r_{ij})}{(\mu_0 r_{ij})^2}-\frac{\sin(\mu_0 r_{ij})}{(\mu_0 r_{ij})^3}\bigg]\bigg\},
\end{multline}
and
\begin{multline}
    M_{ij}=\frac{3}{4}\sqrt{\Lambda_i\Lambda_j}\bigg\{-\bigg[1-(\vec{d}.\vec{r}_{ij})^2\bigg]\frac{\cos(\mu_{0}r_{ij})}{\mu_{0}r_{ij}}\\
    +\bigg[1-3(\vec{d}.\vec{r}_{ij})^2\bigg]\bigg[\frac{\sin(\mu_0 r_{ij})}{(\mu_0 r_{ij})^2}-\frac{\cos(\mu_0 r_{ij})}{(\mu_0 r_{ij})^3}\bigg]\bigg\},
\end{multline}
respectively. For identical atoms $\Lambda_i=\Lambda_j=\Lambda=\frac{\omega^3\mu^2}{3\pi\epsilon \hbar c^3}$. We have  $\mu_0=\omega_0/c$, $r_{ij}=\vert\vec{r_j}-\vec{r_i}\vert$ is the inter-atomic distance, $\vec{r_{ij}}$ is the unit vector along the interatomic axis and $\vec{d}$ the unit vector along the atomic transition dipole moments. $\mu_0 r_{ij}$ set up a length scale into the problem and allows for a discuss of the dynamics in two regimes, a) collective $\mu_0 r_{ij}<< 1$, b) independent  $\mu_0 r_{ij}\geq 1$, decoherence regimes \cite{banerjee2010dynamics}. The collective decoherence regime of this model is used next to investigate the speed limit of the creation and decay of the entanglement by an amount  $E^{\mathcal{B}}(\rho)$ in the two two-level atomic system.  \newline

\begin{figure}
    \centering
    \includegraphics[height=65mm,width=1\columnwidth]{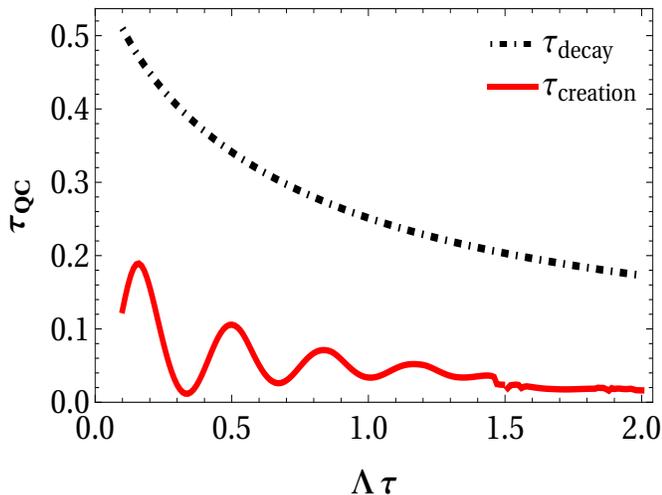}
    \caption{Quantum speed limit time in terms of Bures distance for the decay and creation of quantum entanglement  by an amount $E^{\mathcal{B}}(\rho)$ for a maximally entangled Bell state $\vert\psi^{+}\rangle$ and separable state $\vert g_1 e_2\rangle$ .The coupling parameters are $M_{12}=4.65 \Lambda$ and $\Lambda_{12}=0.95 \Lambda$, with $ \mu_0 r_{12}= 0.08$ indicating the collective decoherence regime. The actual driving time $\tau=1$.}
    \label{qslt_bures_ed_creation_decay}
\end{figure}

We investigate the dynamics of  initially separable ($\vert g_1 e_2\rangle$) and entangled Bell $(\vert\psi^{+}\rangle)$ states.
We show  that $\tau_{QC}$ decreases as the strength of inter-atomic coupling increases. Entanglement is generated in the due course of evolution of  the collective atomic system for an  initial separable state $\vert g_1e_2\rangle$. To estimate the Bures measure of generated entanglement,  the fidelity of entangled state in terms of concurrence (Eq.~\ref{Bures_ent_fidelity}) with the closest separable state is estimated as $F_{P}(\rho)=\frac{1}{2}(1+\sqrt{1-e^{-2 t\Lambda}\vert \sin(2 t M_{12})+i \sinh(t\Lambda_{12})\vert^2})$. It's found that the mixed separable state of the form $\sigma=x\vert00\rangle\langle00\vert+(1-x)\vert11\rangle\langle11\vert$ provides the maximum fidelity for $x=\frac{e^{-t \Lambda} F_{P}(\rho)}{e^{t \Lambda}-\cosh(t\Lambda_{12})}$.  Similarly for the dynamics of initially maximally  entangled  Bell state $(\vert\psi^{+}\rangle)$, we calculate the fidelity of separability as $F_{P}(\rho)=\frac{1}{2}(1+\sqrt{1-e^{-2 t (\Lambda+\Lambda_{12})}})$. In this case, closest separable state takes the form $\sigma=x\vert01\rangle\langle01\vert+(1-x)\vert11\rangle\langle11\vert$ with $x=4 e^{t (\Lambda+\Lambda_{12})} F_{P}(\rho)$. In Fig.~\ref{qslt_bures_ed_creation_decay}, using these time dependent separable states we depict the speed limit time for the creation $(\tau_{\textrm{creation}})$ of quantum entanglement for an initial separable state and the decay $(\tau_{\textrm{decay}})$ of the initial entanglement by an amount  under collective decoherence. From Fig.~\ref{qslt_bures_ed_creation_decay}, for decay of quantum correlations, speed limit time decreases while for the creation of entanglement it is evident that QSL time rises and falls, even though the overall envelope decreases.   \newline
The QSL time for the creation and decay of quantum discord could also be estimated. The explicit analytical expression is available only for specific class of states, for instance Bell diagonal state as mentioned in the previous case. The dynamics of collective atom model does not preserve the Bell diagonal structure of states, which prevents the use of that expression for the calculation of  speed limit of the dynamics of quantum discord.  In general, identifying the closest classical state to estimate the Bures measure of discord for a given state $\rho$ is cumbersome.  We hope to come back to this in the near future.\newline

\section{Conclusions}
We derived the ML-MT type bound on the speed limit time  for the creation and decay of quantum correlations by an amount in a quantum system under the influence of the surrounding  environment. We used Bures distance measure to quantify entanglement and quantum discord. The minimum time required to make a change in these correlations, under the evolution, was estimated.  QSL time for the change in quantum correlations  in terms of operator norm gives a tighter bound for the generation and decay of entanglement and quantum discord. As illustrations of the theoretical developments, we investigated the  speed limit time for the dynamics of QC for specific models. In particular, we considered the OUN dephasing channel and  a dissipative two-atomic system. We showed that QSL time for quantum correlations exhibits a complex behavior as the  strength of the coupling increases.
\section*{Acknowledgement}
SB acknowledges the support from the Interdisciplinary Cyber-Physical Systems (ICPS) programme of the Department of Science and Technology (DST), India, Grant No.: DST/ICPS/QuST/Theme-1/2019/6.

\bibliography{main.bib}

\end{document}